# Microwave Dielectric Heating of Drops in Microfluidic Devices†

David Issadore[a], Katherine J. Humphry[b], Keith A. Brown[a], Lori Sandberg[c], David Weitz[a,b], and Robert M. Westervelt*[a,b]

We present a technique to locally and rapidly heat water drops in microfluidic devices with microwave dielectric heating. Water absorbs microwave power more efficiently than polymers, glass, and oils due to its permanent molecular dipole moment that has a large dielectric loss at GHz frequencies. The relevant heat capacity of the system is a single thermally isolated picoliter drop of water and this enables very fast thermal cycling. We demonstrate microwave dielectric heating in a microfluidic device that integrates a flow-focusing drop maker, drop splitters, and metal electrodes to locally deliver microwave power from an inexpensive, commercially available 3.0 GHz source and amplifier. The temperature of the drops is measured by observing the temperature dependent fluorescence intensity of cadmium selenide nanocrystals suspended in the water drops. We demonstrate characteristic heating times as short as 15 ms to steady-state temperatures as large as 30°C above the base temperature of the microfluidic device. Many common biological and chemical applications require rapid and local control of temperature, such as PCR amplification of DNA, and can benefit from this new technique.

## Introduction

The miniaturization of the handling of liquid and biological samples has enabled great advances in fields such as drug discovery, genetic sequencing and synthesis, cell sorting, single cell gene expression studies, and low-cost portable medicine.[1-7] At the forefront of this technology are the micro-fabricated pipes, valves, pumps, and mixers of microfluidics that are leading to integrated lab-on-a-chip devices which are causing a paradigm shift in fluid handling analogous to what integrated circuit technology did for electronics half of a century ago.[6] A growing library of elements for lab-on-a-chip systems have been developed in recent years for tasks such as the mixing of reagents, detecting and counting of cells, sorting cells, genetic analysis, and protein detection.[1-7] There is one function, however, that is crucial to many applications and which has remained a challenge: the local control of temperature. The large surface area to volume ratios found in micrometer scale channels and the close proximity of microfluidic elements make temperature control in such systems a unique challenge.[8,9]

Much work has been done in the last decade to improve local temperature control in microfluidic systems. The most common technique uses Joule heated metal wires and thin films to conduct heat into fluid channels.[10-14] In these devices the thermal conductivity inside the microfluidic devices control the localization of the temperature change and tends to be on the order of centimeters.[9,11] Temporal control is limited by the heat capacity and thermal coupling of the microfluidic device to the environment and thermal relaxation times tend to be on the order of seconds.[9,11] Alternative techniques to improve the localization and response time have been developed, such as those that use integrated micrometer size Peltier junctions to transfer heat between two channels containing fluid at different temperatures.[15] Fluids may also be cooled on millisecond time scales with evaporative cooling by pumping gasses into the fluid channels.[16]

The focus of this research is to integrate electronics with microfluidics to bring new capabilities to lab-on-a-chip systems.[6,7] In this paper we present a technique to locally and rapidly heat water in drop based microfluidic systems with microwave dielectric heating. The devices are fabricated using soft lithography and are connected to inexpensive commercially available microwave electronics. Because the drops are thermally isolated from the bulk of the device exceptionally fast thermal cycling times $\tau_s$ = 15ms are attained, raising the drops' temperature to 30ºC above the base temperature of the device.

Rapid, localized heating with microwave dielectric heating in microfluidic devices has exciting applications in many common biological, chemical, and medical applications. The amplification of DNA using polymerase chain reaction (PCR) is an application that has gotten a lot of attention.[10-16] PCR requires thermal cycling between 65, 85, and 95°C. Our device is an especially good fit for droplet based PCR, which can be used to rapidly test large assays of small volume samples.

### Theory of Dielectric Heating

Dielectric heating describes the phenomenon by which a material is heated with a time-varying electric field. Induced and intrinsic dipole moments within an object will align themselves with a time-varying electric field. The energy associated with this alignment is viscously dissipated as heat into the surrounding solution. The power density P absorbed by a dielectric material is given by the frequency ω of the applied electric field, the loss factor ε'' of the material, the vacuum permittivity $\varepsilon_0$, and the electric field strength |E| with the expression:

[a] School of Engineering and Applied Sciences, Harvard University, Cambridge, MA 02138, USA
[b] Department of Physics, Harvard University, Cambridge, MA 02138, USA
[c] College of Engineering and Applied Sciences, University of Wyoming, Laramie, WY 82071, USA
* E-mail: westervelt@seas.harvard.edu

$$P = \omega \varepsilon_o \varepsilon'' |E|^2 \quad (1)$$

The loss factor of the material is dependent on the frequency of the electric field and the characteristic time τ of the dielectric relaxation of the material, with the expression:

$$\varepsilon'' = \frac{(\varepsilon_s - \varepsilon_\infty)\omega\tau}{1 + (\omega\tau)^2} \quad (2)$$

where $\varepsilon_s = 78.4\varepsilon_0$ is the low frequency dielectric constant of water and $\varepsilon_\infty = 1.78\ \varepsilon_0$ is the optical dielectric constant.

Due to water's large dielectric loss at GHz frequencies microwave power is absorbed much more strongly by water rather than PDMS or glass. Power is most efficiently absorbed by water with microwaves that have a frequency that coincides with the relaxation time of water τ = 9 ps which corresponds to approximately an 18 GHz signal.[17] Our device operates at 3.0 GHz, a frequency very close to that of commercial microwave ovens (2.45 GHz), where water still readily absorbs power. It is inexpensive to engineer electronics to produce and deliver 3.0 GHz frequencies because it is near the well-developed frequencies of the telecommunications industry.

### Model of Dielectric Heating of Drops

Two independent figures of merit describe the heater, the steady-state change in temperature ΔT that the drops attain and the characteristic time $\tau_s$ that it takes to change temperature. The steady-state temperature occurs when the microwave power entering the drop equals the rate that heat leaves the drop into the thermal bath. The thermal relaxation time depends only on the geometry and the thermal properties of the drops and the channel and is independent of the microwave power.

To describe our heater we use a simplified model in which the temperature of the channel walls do not change. The thermal conductivity of the glass and PDMS channel walls is much larger than that of the fluorocarbon(FC) oil in which the drops are suspended, which allow the glass and PDMS mold to act as a large thermal reservoir that keep the channel walls pinned to the base temperature.

The drop is modelled as having a heat capacity that connects to the thermal reservoir through a thermal resistance. The drop has a heat capacity $C = VC_w$ that connects to the thermal reservoir with a thermal resistance $R = L/Ak_{oil}$, where V is the volume and A is the surface area of the drop, $C_w$ is the heat capacity per volume of water, L is the characteristic length between the drop and the channel wall, and $k_{oil}$ is the thermal conductivity of the oil surrounding the drop. A steady-state temperature ΔT is reached when the microwave power PV entering the drop is equal to the power leaving the drop $\Delta T k_{oil} A/L$. We find the steady-state temperature ΔT = PVR to be:

$$\Delta T = \frac{V}{A}\frac{L}{k_{oil}}P \quad (3)$$

The system has a characteristic time scale $\tau_s$ = RC that describes the thermal response time of the system:

$$\tau_s = \frac{V}{A}\frac{L}{k_{oil}}C_w \quad (4)$$

This simplified model describes several key features of the microwave heater. The steady-state temperature is linearly proportional to the microwave power, whereas the characteristic thermal relaxation time is independent of the microwave power. The characteristic time and the steady-state temperature are both proportional to the volume to surface ratio of the drops. A trade-off relation exists between the rate of heating $1/\tau_s$ and the steady-state temperature, whereby a larger volume to surface ratio increases the speed of the heater but decreases its steady-state temperature for a given microwave power, and vice versa. Similarly an increase in the ratio of the characteristic length between the drop and the channel wall and the thermal conductivity of the oil $L/k_{oil}$ decreases the rate of heating and increases the steady-state temperature.

### Methods

The devices are fabricated using poly(dimethylsiloxane) (PDMS)-on-glass drop-based microfluidics. A schematic cross-section of the device is shown in Fig. 1a. Microwave power is locally delivered via metal electrodes that are directly integrated into the microfluidic device and that run parallel to the fluid channel. The drops are thermally insulated from the bulk by being suspended in low thermal conductivity oil.

Drops are created using a flow-focusing geometry[18] as is shown in Fig. 2a. A fluorocarbon oil (Fluorinert FC-40, 3M) is used as the continous phase and the resulting drops contain 20 μM of dissolved carboxyl coated CdSe nanocrystal (Invitrogen) suspended in phosphate buffered saline (PBS) solution. A surfactant comprised of a polyethyleneglycol (PEG) head group and a fluorocarbon tail (RainDance Technologies) is used to stabilize the drops.[19] The walls of the microfluidic channels are coated with *Aquapel®* (PPG Industries) to ensure that they are preferentially wet by the fluorocarbon oil. Fluid flow is controlled via syringe pumps. Typically, drops made in our flow-focussing geometry have a diameter of 50 μm. To ensure an appropriate volume fraction of drops, the oil flow rate is ~10x that of the aqueous phase.

To make drops smaller than the channel height, and thus separated from the walls of the channel to ensure adequate thermal isolation, we use drop splitters[20] as is shown in Fig. 2b. The drop splitters are designed to break each drop into two drops of equal volume. Passing our drops through two drop splitters in series decreases the radius of our drops by a factor of $(½)^{⅔}$ = 0.63. Drop splitters allow the device to be made in a single fabrication step, because they remove the necessity of making the drop maker with a channel height smaller than the rest of the device.[18]

The metal electrodes are directly integrated into the PDMS device using a low-melt solder fill technique.[21] The masks for the soft lithography process are designed to include channels for fluid flow and a set to be filled with metal to form electodes. After inlet holes have been punched into the PDMS and the PDMS is bonded to a glass slide, the microfluidic device is placed on a hot plate set to 80°C. A 0.02 inch diameter indium alloy wire (Indalloy 19; 52% Indium, 32.5%

Bismuth, 16.5% Tin from Indium Corporation) is inserted into the electrode channel inlet holes and, as the wire melts, the electrode channels fill with metal via capillary action. The resulting electode channels run along either side of the fluid as is shown in Fig. 2c. To keep the drops from heating from the fringe electric fields before the drop enters the heater, the fluid drops are pressed against the PDMS wall which acts as a thermal reservoir by narrowing the channel to 20 μm wide.

The electronics that create the microwave power are assembled using inexpensive modules. The microwaves are generated with a voltage controlled oscillator (ZX95-3146-S+, Mini-Circuits) and amplified to a maximum of 11.7 V peak-to-peak with a maximum power of 26dBm with a power amplifier (ZRL-3500+, Mini-Circuits). The microwave amplifier connects with a cable to a sub miniature assembly (SMA) connector mounted next to the microwave device. Copper wires approximately 2 mm in length connect the SMA connector to the metal electrodes in the PDMS device. Our electronics operate at 3.0 GHz where water's microwave power absorption is roughly 1/3 as efficient as at water's optimum absorption frequency (~18 GHz) but where electronics are inexpensive and commercially available. The electronics used in our system costs less than $US200 and are easy to setup.

Finite element simulations are performed to determine the electric field strength in the microwave heater which is used to calculate the microwave power absorbed by the drops. Fig. 1b shows a schematic cross-section of the microfluidic device where the drops pass between the metal electrodes. The channel cross-section has dimensions 50 x 50 $\mu m^2$. Parallel to and 20 μm away from each side of the fluid channel are metal lines that are 100 μm wide and 50 μm high. Superimposed on the schematic in Fig. 1 is a quasi-static electric field simulation of the electric field (Maxwell). We find that for a 12 V peak-to-peak signal across the metal lines, the RMS electric field within a drop with a 15 μm radius suspended in fluorocarbon oil is $|E| \sim 8\times10^3$ V/m. The electric field linearly scales with the voltage across the metal lines which allows us to calculate the field within the drop for any voltage. The simulated electric field is combined with Eq. 1 and Eq. 2 to calculate the microwave power that enters the drops which may be combined with Eq. 3 to predict steady-state temperatures.

The temperature of the drops is measured remotely by observing the temperature-dependent fluorescence of CdSe nanocrystals embedded in the drops. To calibrate our thermometer, we turn the microwave power off and use a hot plate to set the temperature of the microfluidic device. The fluid channel is filled with CdSe nanocrystal suspended in water and the temperature of the hot plate is slowly increased from 25°C to 58°C while the fluorescence intensity of the CdSe quantum dots is measured. The measured fluorescence intensity is plotted as a function of temperature in Fig. 1c. We fit a line with slope 0.69 %/°C ± 0.03 %/°C to the data and use this relationship to convert fluorcence measurements into temperatures. The device is monitored with an Hamamatsu ORCA-ER cooled CCD camera attached to a BX-52 Olympus microscope. Images are taken with MicroSuite Basic Edition by Olympus and analysed in MATLAB (The MathWorks, Inc.). The microfluidic device is connected with an SMA to the microwave amplifier and sits on top of a hot plate underneath the microscope as is shown in Fig. 2d.

We test our devices by measuring the temperature change of water drops as they travel through the microwave heater. A long-exposure fluorescence image of many drops traveling through the microwave heater shows the ensemble average of the temperature change of drops at each point in the channel. A plot of the drop heating in time may be extracted from this image using the measured flow rate of the drops through the microfluidic system. An experiment is performed with the flow rate of the oil at 15μL/hr and the water at 165μL/hr. A bright field, short shutter speed image is taken of the drops traveling through the microwave heater and the drops' average diameter is measured to be 35μm. The microwave heater is turned on with a frequency of 3.0 GHz and a peak to peak voltage of 11V. A long exposure (2 seconds) fluorescent image is taken of the microwave heater that is normalized against images taken with the microwaves turned off to remove artificats that arise from irregularities in the geometry of the channel, the light source, and the camera.

## Results

The drops are heated to a steady-state temperature ΔT as they pass through the microwave heater. Figure 3a shows the normalized fluorescence intensity of the drops as they enter the microwave heater superimposed onto a bright-field image of the device. It can be seen that as the drops enter the channel their average fluorescent intensity drops which shows that they are being heated. A line average of the normalized image is taken in the direction perpindicular to the fluid flow and is plotted against the length of the channel, as in Fig. 3b. As the drops enter the channel their average fluorescent intensity falls exponentially to 85% of its initial intensity in 300 μm.

The drops are heated to steady-state temperatures ΔT as large as 30ºC above the base temperature of the microfluidic device in only $\tau_s$ = 15 ms. The avarage temperature change of the drops as a function of time is plotted in Fig. 3c. For this heating power, the temperature rises to a steady-state value of 26°C above the base temperature of 21°C in 15ms. We arive at the curve by using the calibration curve of the CdSe nanocrystals, Fig. 1c, to convert the fluorescent intensity in Fig. 3b into temperature. The sum of the flow rates of the oil and water are used to calculate the speed of the drops through the channel, which may be used to transform the length in Fig. 3a into the time that the drops have spent in the heater. Each data point consists of the average of 20 independently taken 2 sec exposures. The combined statistical error of the measurement of the heating is also plotted in Fig. 3c. The average error is ~1.4°C.

The steady-state temperature change ΔT of the drops may be set from 0ºC to 30ºC by varying the applied microwave power as is described in Eq. 3. The microwave power is controlled by experimentally varying the peak-to-peak voltage of the applied microwave voltage which varies the strength of the electric field inside the drops as is described by our simulations. A series of plots of temperature versus time for different applied powers is shown in the insent of Fig. 4a, and shows steady-state temperature changes ranging from 2.8°C to 30.1°C, with an average error of 1.5°C . It is noteworthy that all of the heating curves have an exponential form and have the same characteristic rise time $\tau_s$ = 15 ms.

We compare the steady-state temperatures observed in Fig. 4a for different applied microwave powers with our model of microwave heating and find good agreement. The steady-state temperature change is plotted versus the microwave power density in Fig. 4a and is fit with a line. As is expected from Eq.3 the steady-state temperature rises linearly with applied microwave power. We calculate the microwave power density using the electric field values determined from simulations (Fig. 1b). The only variable in Eq. 3 that we do not measure or that is not a material property is L the characteristic length scale between the drop and the channel wall. We estimate this characteristic length L = 28μm using the measured steady-state temperatures (Fig. 4a) and the known material properties using Eq. 3.

We compare the observation that the temperature approaches steady-state exponentially in time in Fig. 4b with our model of microwave heating and find good agreement. To compare the model's prediction that the drops approach equilibrium exponentially in time with a single relaxation time constant (Eq. 4) with our observations, we plot the temperature change versus time on a semi-log plot and fit with a line. As is predicted by our model the drops approach equilibrium exponentially and with a single time constant. The thermal relaxation time constant is measured to be $\tau = 14.7 \pm 0.6$ ms. The only variable in Eq. 4 that we do not measure is the characteristic length L between the drop and the channel wall. We estimate this characteristic length L = 35 μm using the measured characteristic time constant and known material properties using Eq. 4. The independent measurements of the thermal relaxation time and the steady-state temperature versus power give two independent measures of the length L that are within 20% of each other. The agreement between the two independent measurements supports our simple model for microwave heating of drops.

## Discussion

We have demonstrated an integrated microfluidic microwave dielectric heater that locally and rapidly increases the temperature of drops of water in oil. The large absorption of microwave power by water relative to oil, glass, and PDMS allows local and rapid heating in microfluidic devices without difficult fabrication. Both improving the insulation of drops from the channel walls and increasing the volume to surface area ratio of the drops would allow for larger temperature changes.

Microwave dielectric heating of drops is well suited for integration with hybrid integrated circuit (IC) / microfluidic systems.[6,7] Dielectrophoresis chips use arrays of 11 μm x 11 μm electrodes to trap and move drops or cells inside a microfluidic chamber[7]. By using GHz frequencies, one can use the same chip to locally heat a single drop. Chip based PCR would be would prove to be a valuable tool for DNA analysis[6].

Microwave dielectric heating has many exciting scientific and technological applications. One noteworthy application for rapid, localized heating in microfluidics devices is PCR.[11-16] Our heater is uniquely suited for drop-based PCR, allowing for the rapid amplification and analysis of large populations genes and enzymes.The rapid heating acheived with our technique might also be used to set temperatures rapidly and controllably in biological and chemical assays where observations are made on the millisecond time scale.[23]

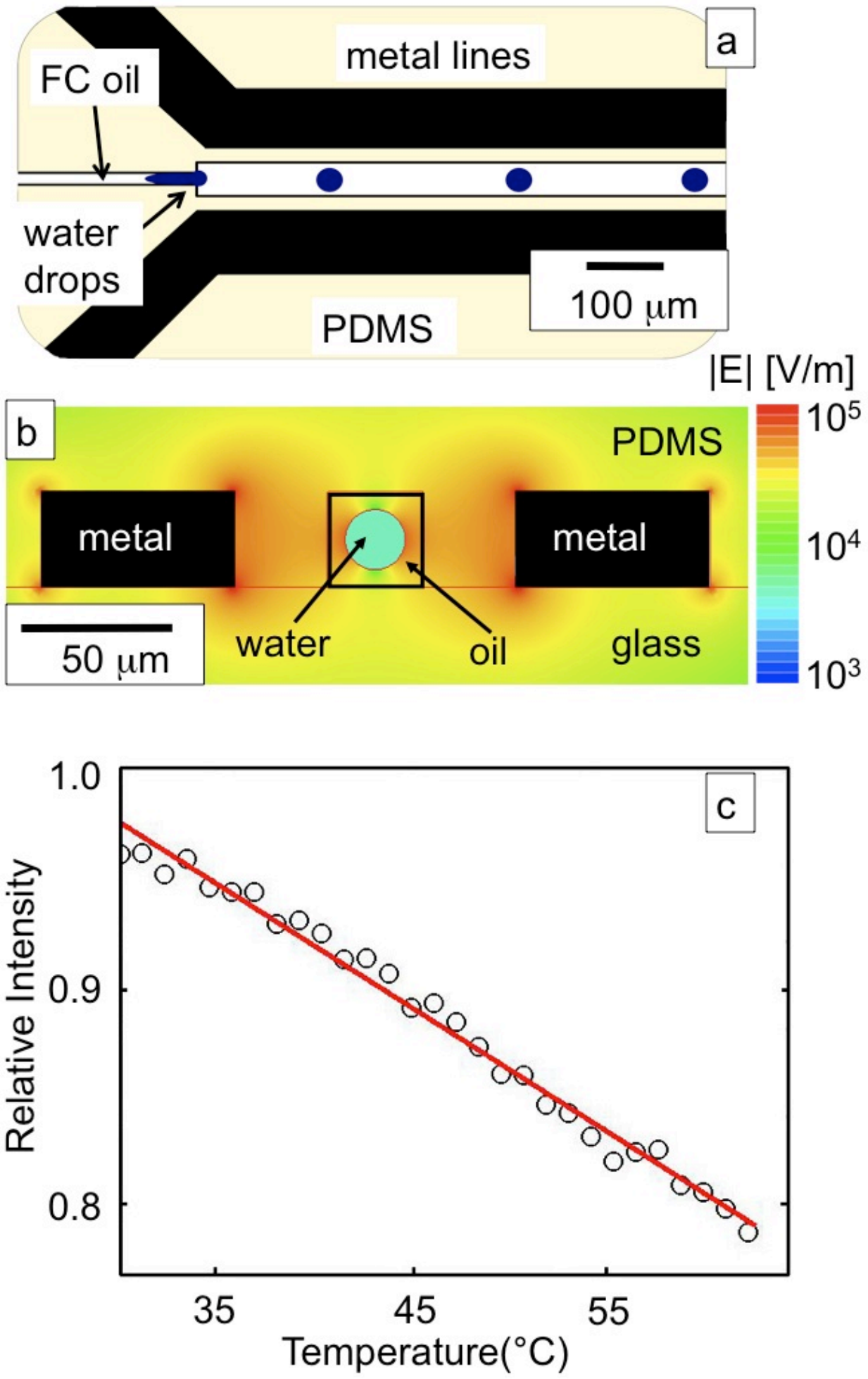

**Fig. 1** (a) A schematic of the microwave heater. The black metal lines represent the metal lines which are connected to the microwave source, the center fluid channel carries drops of water immersed in fluorocarbon (FC) oil, (b) a cross section of the microwave heater with a quasi-static electric field simulation superimposed is shown, the electric field is plotted in log scale, (c) The calibration curve of the CdSe nanoparticles which are used as temperature sensors, where the blue circles are data points and the red line is the fit.

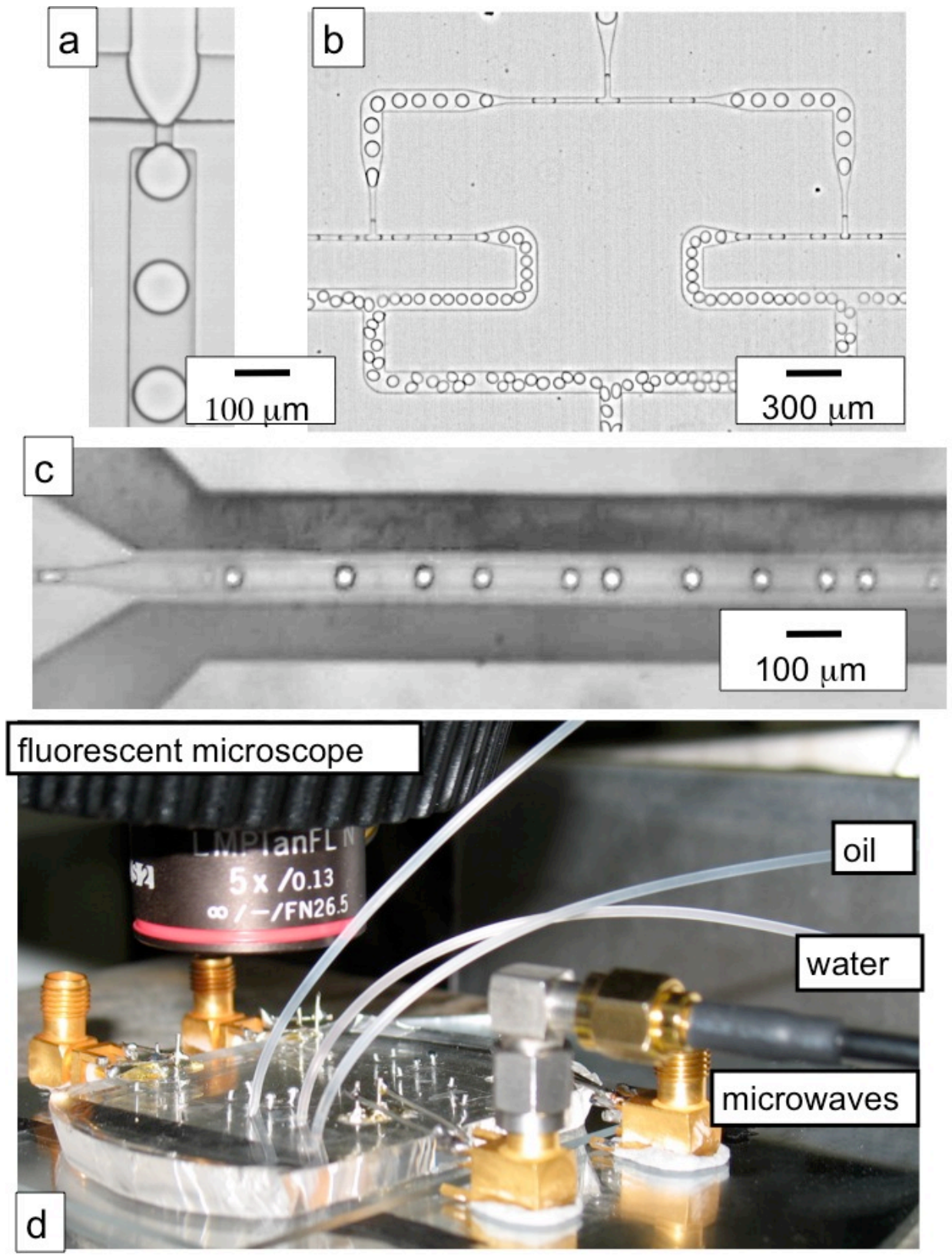

**Fig. 2** (a) A micrograph of the flow-focusing drop maker, (b) The two sets of drop splitters in series, (c) a micrograph of the microwave heater, the dark regions that run parallel to the fluid channel are the metal lines, (d) a photograph of the microfluidic device, connected to the microwave amplifier with an SMA cable, on top of a hot place, underneath the fluorescent microscope.

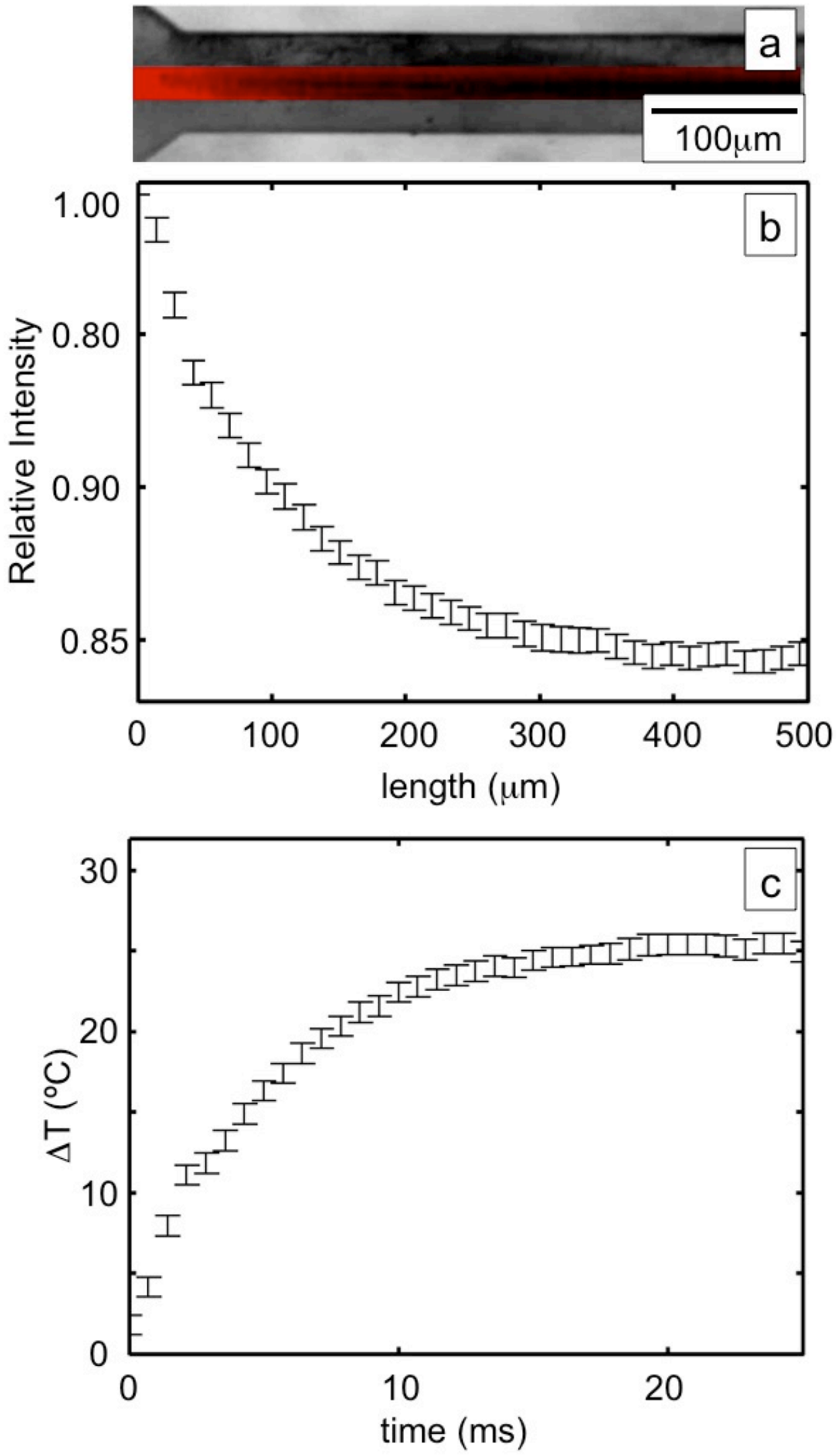

**Fig 3.** (a) A 2 second exposure of the fluorescent signal normalized to an image taken with the microwave source turned off, (b) the red curve shows a line average of the normalized intensity plotted versus distance down the channel. The green curve shows the temperature plotted versus time of the drops heating.

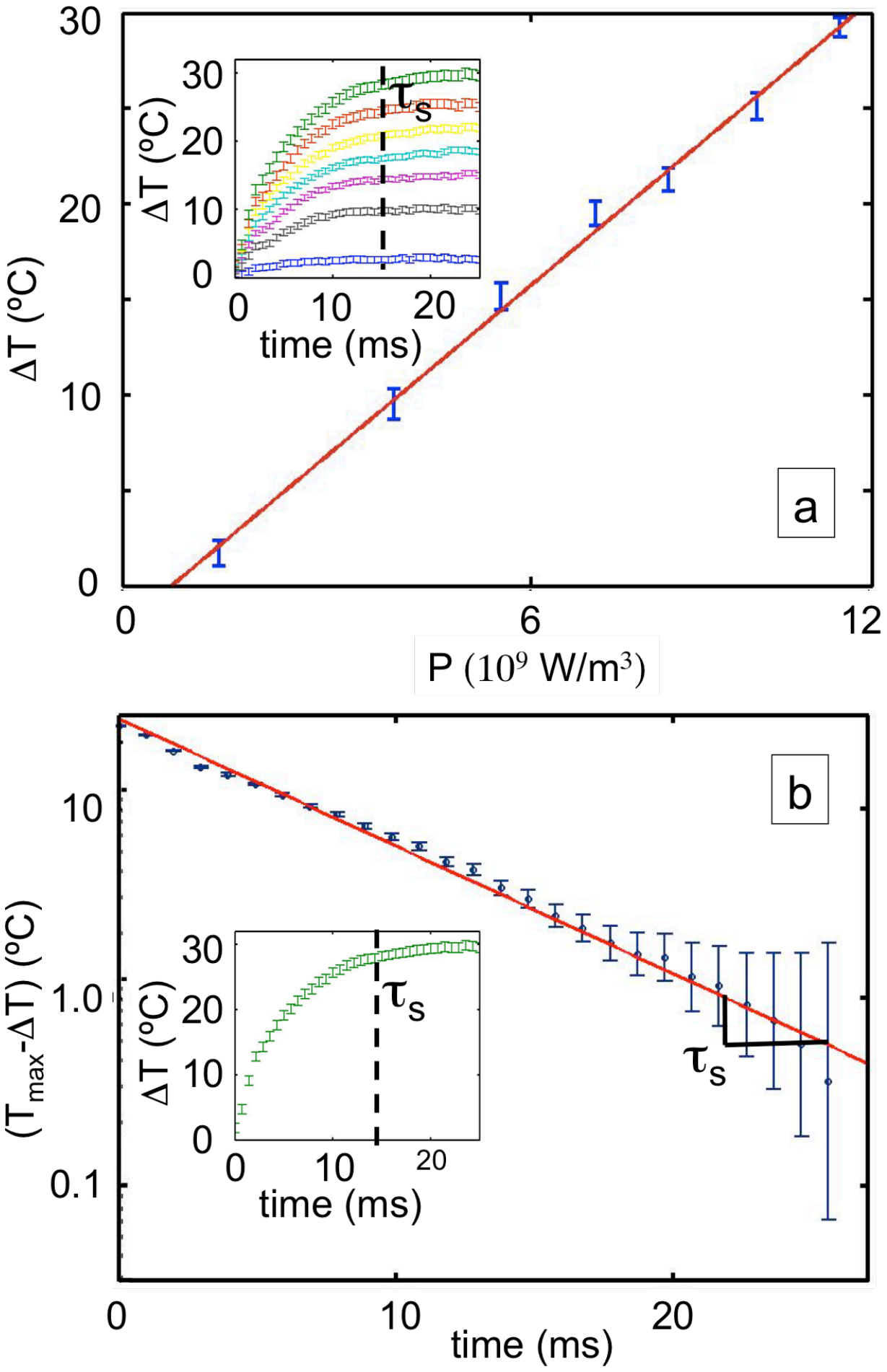

**Fig 4.** (a) The steady-state temperature of the drops is controlled by varying the amplitude of the microwave voltage. In the inset the green curve is at 11.7V, the red 11.0V, the yellow 10.3V, the light blue 9.3V, the purple 8.5V, the grey 7.5V, the blue 4.5V, (b) the steady-state temperature is plotted versus the power density, as calculated by Eq. 1. The magnitude of the change in temperature versus time plotted on a log-linear scale. The temperature rises as a single exponential with a characteristic time, $\tau_s$= 14.7 ± 0.56 ms.